# Antenna design for the SKA1-LOW and HERA super radio telescopes

Eloy de Lera Acedo
Cavendish Astrophysics
University of Cambridge
Cambridge, UK
eloy@mrao.cam.ac.uk

Hardie Pienaar
Cavendish Astrophysics
University of Cambridge
Cambridge, UK

Nicolas Fagnoni
Cavendish Astrophysics
University of Cambridge
Cambridge, UK

*Abstract*—This paper summarizes the design process and metrics for the latest antenna design for 2 radio telescopes, SKALA4 for the SKA1-LOW instrument and the V-feed for the HERA telescope. In the paper we briefly describe the main features of the antenna element design and the most important figures of merit for both instruments. Finally, we show the response of both designs against some of these figures of merit.

## I. INTRODUCTION

The Square Kilometre Array (SKA) [1] and the Hydrogen Epoch of Re-ionization Array (HERA) [2] are two of the most prominent, technically advanced large N (antenna number) aperture arrays as well as ambitious radio astronomy projects in development today. These two projects will count with a large number of aperture antennas (512 stations with 256 antenna elements each in the case of the SKA1-LOW instrument and 350 14-m dish antennas in the case of HERA). SKA1-LOW (the low-frequency instrument of SKA) and HERA have several common aspects as well as differences in both their scientific aims and its design goals and requirements, and so do their array antennas.

The SKA represents the largest and most powerful radio telescope at meter and centimeter wavelengths that has ever existed. Its extraordinary sensitivity and flexible design will allow SKA tackle a large number of key scientific projects. The SKA1-low instrument (50-350 MHz) will populate the semi-deserted areas of Western Australia and will carry out detailed studies of the very early Universe, the birth of the first starts and its evolution. Furthermore, the SKA1-LOW will also research on Pulsars and other transients as well as a series of other exciting scientific cases. SKA's all digital approach with more than 131,000 antenna elements electronically scanning the array in every direction calling for an array antenna design that can provide the demanding requirements of SKA: sensitivity, field of view, spectral smoothness, polarization purity, etc in an ultra-wide frequency band (7:1). Aspects such as the matching the antennas to the low noise amplifiers are key for both achieving superb sensitivity as well as spectral smoothness. The mutual coupling in the stations is also a key component in the design and especially the calibration of these antennas. Furthermore, the SKA numbers call for a robust, low-cost design to be produced in high volume.

HERA on the other hand is a focused experiment targeting the detection of the power spectrum signature signal from the Epoch of Re-ionization (EoR). It is also a wideband instrument (50-250 MHz) that will use the so-called Delay Spectrum technique for the analysis of the received signal. This technique calls for an antenna design with a time-domain transfer function as close to the delta function as possible (a truly wideband design). With a much less demanding cost budget and narrower field of view, HERA is currently being installed in the Karoo radio reserve, South Africa.

In this paper we give a brief summary of the design challenges and design process for the antennas of these two telescopes. This design has been led by researchers at the Cavendish Laboratory, University of Cambridge in collaboration with Cambridge Consultants Ltd. The antenna design work is part of a design consortium in the case of SKA (The Aperture Array Design Consortium) and a design team (the analogue team) in the case of HERA. The SKA antenna is a log-periodic antenna currently in its 4th iteration (SKALA4) after a selection process in the summer of 2017 and the HERA feed is a Vivaldi type antenna. In section II we discuss the different antenna metrics and we explain what makes both projects different in terms of the requiemrents for antennas. Section III (SKALA4) and IV (HERA V-feed) are then dedicated to present the main aspects of the antenna design and their response against some of the aforementioned metrics. Finally we draw some conclusions in setion V.

## II. ANTENNA METRICS

The antenna design for SKA1-LOW and HERA share similar engineering requirements as well as some that are different. These differences are driven by the specific techniques these telescopes are expecting to use as well as by the size of the projects and the choice of technology (eg. array antennas on top of a ground mesh (SKA) vs array of dishes (HERA)). In the follwing lines we describe these antena metrics.

- *Sensitivity*. Defined as the effective area over the system temperature (A/T) [3]. This requirement has driven the design of the SKA1-LOW antenna since the beginning of the design process [4]. The SKA will rely in its super sensitivity to tackle a broad range of scientific questions and high sensitivity is key to achieve the required level of calibration for the system [4]. HERA also requires high sensitivity to detect the ~mK level signal from the Epoch of Re-ionization, but focused on the use of the so-called foreground avoidance technique [5], the antenna response in the delay spectrum is more critical than sensitivity. In both cases the sensitivity is strongly dominated by the directivity within the field of view of the instrument and the noise matching to the first Low Noise Amplifier (LNA), especially at the high end of their frequency range, where the sky noise (mostly from Galactic foregrounds) has decreased below the receiver temperature.

- *Spectral smoothness*. This is a critical metric for both instruments. Tackling a very broad range of science cases SKA1-LOW requires frequency smoothness at different levels (large frequency scales as well as small frequency scales). Frequency smoothness is especially important on SKA for its calibration, requiring smoothness and stability in a large set of parameters (sensitivity, power pass-band, polarization, etc). HERA however focuses on minimizing the antenna response in the delay spectrum [5]. This can be simply described as the transfer function of the antenna in time domain [6]. HERA requires a clean response in the delay spectrum within the delays of interest (delays larger than ~60ns) in order not to contaminate the so-called EoR window of observation (set of spatial and spectral modes where HERA is attempting to detect the EoR). This metric is heavily dominated by the power matching between the antenna and the LNA [7] as well as by reflections on the system (reflections between dish and feed in the case of HERA, mutual coupling, cables, etc).
- *Polarization*. In both cases the polarization purity of the antennas is important, since our knowledge of the polarization state of the foreground signals is limited. For SKA, the way to measure the polarization response of the antennas has been focused around its calibratability (eg. by using the intrinsic cross polarization ratio [8]) while on HERA the work has focused on understanding how the foreground polarized components may affect the EoR window [9].
- *Cost*. This is one of the key differences between both projects. In the case of SKA1-LOW (131,000 elements) the cost constrains are a lot tighter than on HERA. Furthermore, while they are in very similar environments, SKA is expected to operate for several decades while HERA is expected to work for a few years only. This makes the design of the antenna mechanics a crucial and integral part of the whole antenna design for SKA1-LOW while in the case of HERA it has been a follow on process after the electromagnetic design was realized. This, together with the size of the project (SKA is considerably larger as an international project in budget, number of partners involved, etc) has resulted in a very different time scale for the development of the antenna design between both projects (close to a decade for SKA vs ~2 years for HERA).
- *Logistics/maintenance/deployability*. A large part of the antenna mechanical design depends on the requirements for maintainability and deployability. This is again a case where there is a substantial difference between both projects, due to the large number of antennas in SKA1-LOW.

Some other differences between both antenna designs include:

- *Antenna location.* In the case of SKA1-LOW the antennas are arranged in pseudo random arrays with 256 elements and 38m in diameter that imposes a more challenging coupling environment. On the other hand, the pseudo randomize positions tend to randomize the effects of mutual coupling. On the other hand, the HERA feed is located at the focal point of a non-steerable (pointing at zenith) 14 m dish. The shorter baseline to the nearest antenna is therefore considerably larger than for SKA but however the antennas are located in a regular triangular grid. While understanding the effects of mutual coupling will be absolutely crucial for the operation of both instruments, the main antenna design has been mostly driven by the design of the individual element in the absence of strong mutual coupling effects.
- *The effect of the ground.* While the SKA1-LOW antenna is located directly on top of a large ground mesh, the HERA feed is at the focal point of a non-steerable 14-m dish. The presence of metallic reflectors around these antennas has been an important consideration in both cases since it tends to have a negative effect in most of the electromagnetic metrics.

TABLE I. SUMMARY OF METRICS AND DIFFERENCES AFFECTING THE ANTENNA DESIGN PROCESS IN SKA1-LOW AND HERA

| Parameter | *SKA1-LOW* | *HERA* |
|---|---|---|
| Type of instrument/experiment | Ultra large observatory with multiple science cases | Mid-scale experiment |
| Sensitivity | It is most important: Directivity, Trec (LNA) | Delay response is most important: Reflections, matching |
| Spectral smoothnes | In A/T, in power gain. Wide range of science cases and calibration is complex. | Spectral smoothness: Delay response. |
| Polarization | IXR, beam accuracy | Leakage in the Delay Spectrum |
| Cost/mechanics | Cost: Ultra low cost (131,000 antennas) | Cost: Less critical (price per antenna) |
| Antenna location in the array | Mutual coupling is randomized | Mutual coupling is lower but not randomized |
| Other important aspects of the design | Maintenance, deployment time: critical. | Maintenance and deployment requirements are important but more relaxed ("only" 350 dishes) |

III. SKALA4

The latest itereation of the SKA1-LOW antenna, SKALA4 (see Fig. 1) is a log-periodic antenna with 16 dipoles and 2.1 m tall. The antenna's bottom dipole is 1.6 m across producing a square footprint 1.13 x 1.13 m on the ground. The SKALA4 design targeted imporvements over previous iterations [3, 10, 11] on spectral smoothness, sensitivity and polrization response. A full paper will descrive this design in detail. In the current paper we show some of the improvements realized with SKALA4.

Figure 2 shows the predicted sensitivity at zenith of SKA1-Low using SKALA-4, 3 and 2. The imporvement in smoothness as well as sensivity level is evident. These has been realized by adding more dipoles (16 vs 9) and placing them closer together as well as increasing the antenna height. This change improved the directivity and effective area across the field of view of interest (+/-45 degrees from zenith) at the expense of a reduction of sensitivity towards the horizon. Further imporvements were obtained by tuning the dipole shape and the LNA input matching network.

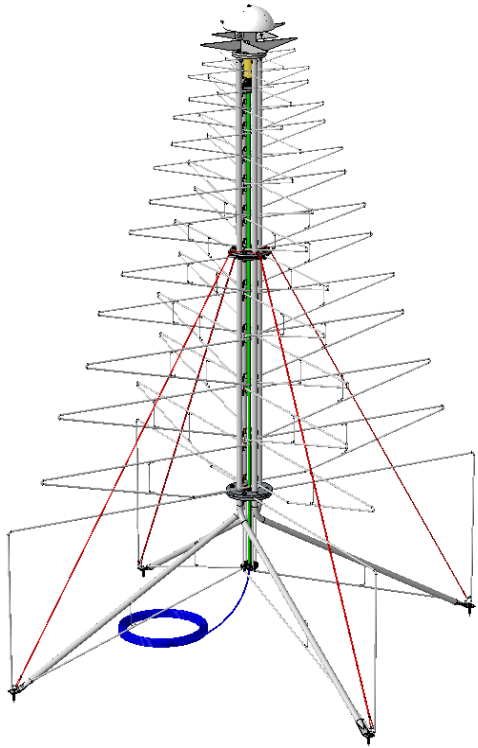

Fig. 1. SKALA4 antenna (computer model)

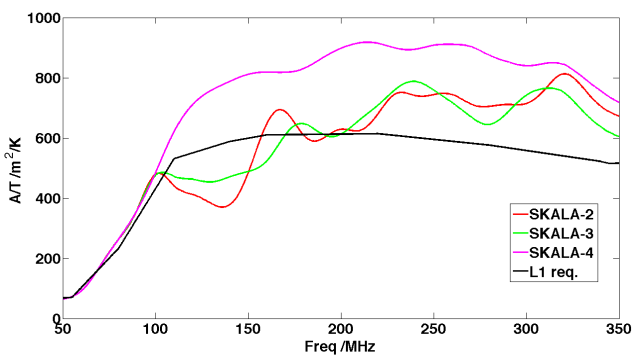

Fig. 2. SKA1-LOW predicted sensitivity

## IV. HERA V-FEED

The design of the new HERA feed was driven by an aim of enlarging the current frequency band (100-200 MHz) to 50-250 MHz. This will allow HERA to realize observations at longer wavelengths (where information about the birth and evolution of the first starts – Cosmic Dawn- is expected to be encoded) and also shorter wavelengths (where astronomers expect to study the end of the EoR). A full paper will describe in detail the feed design.

With this goal, a Vivaldi antenna was designed. The choice of a Vivaldi antenna was driven by the aim of avoiding the need of a back plane for the feed that could result on reflections in band degrading the spectral smoothness of the antenna. Figure 3 shows the typical beam of a Vivaldi antenna. In principle, this type of antenna would not need a backing structure to direct its beam towards the dish and therefore could have a good chance to optimize the delay response.

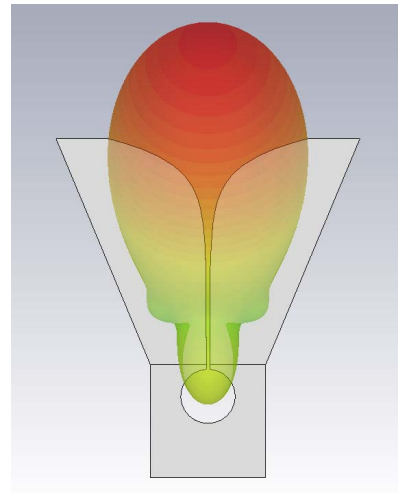

Fig. 3. Vivaldi antenna. Beam at 150 MHz.

Figure 4 shows a view of the mechanical design of the V-feed for HERA including the triangular support frame used to support the antenna from 3 poles and locate it at the focal point of the dish. The mechanical design has been realized in collaboration with Cambridge Consultants Ltd.

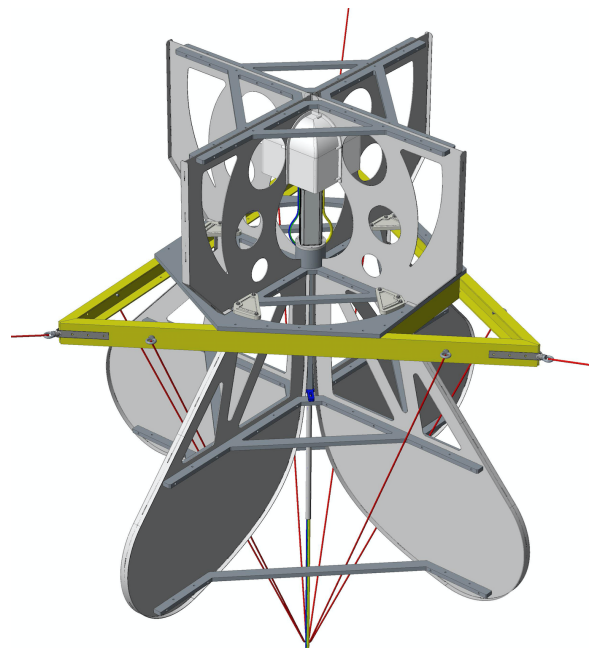

Fig. 4. HERA V-feed (computer model)

Figures 5 and 6 show the simulated response of the HERA dish including both the current cross-dipole feed and the new V-feed to demonstrate the expected improvement across a wider frequency range in realized gain and in the delay spectrum response.

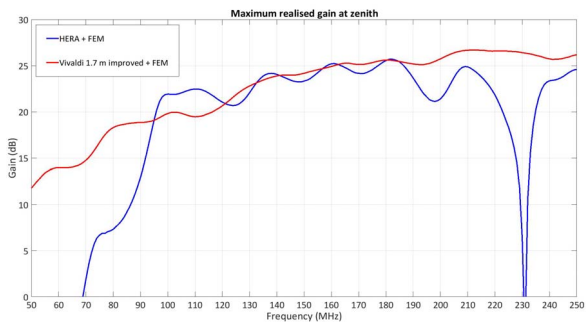

Fig. 5. HERA V-feed vs old dipole feed (realized gain at zenith)

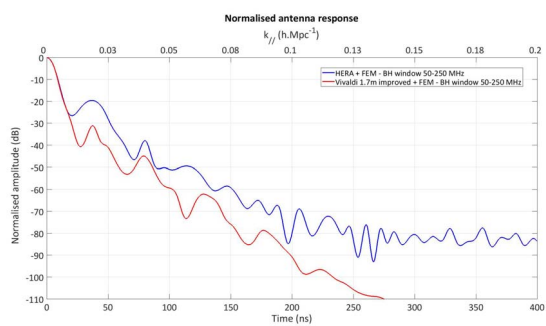

Fig. 6. HERA V-feed vs old dipole feed (Delay response)

## V. CONCLUSIONS

Antenna design for large aperture arrays such as the SKA1-LOW instrument and the HERA telescope requires a comprehensive balance of metrics, both functional and non-functional. The latest iterations of the antenna design for these 2 instruments (SKALA4 and the V-feed) are currently being tested before full deployment starts.

This paper summarizes the main antenna design considerations and metrics for both instruments highlighting the differences and showing simulated analysis of the expected performance of the telescopes using the latest antenna designs.

### ACKNOWLEDGMENT

The authors would like to thank their SKA and HERA coleagues for all the support work and valuable discussions and interactions on the subject. Especifically, the authors would like to thank the LFAA Antenna and LNA WP contributors and the HERA analogue team. This paper summarizes work contributed by both teams. Furthermore, the authors would like to acknowledge STFC for their financial support for this research work.

### REFERENCES

[1] The Square Kilometre Array, www.skatelescope.org
[2] The Hydrogen Epoch of Re-ionization Array, http://reionization.org
[3] De Lera Acedo, E., Razavi-Ghods, N., Troop, N., Drought, N., & Faulkner, A. J. "SKALA, a log-periodic array antenna for the SKA-low instrument: design, simulations, tests and system considerations", Experimental Astronomy, 39(3), 567-594. doi:10.1007/s10686-015-9439-0.
[4] Turner W., 2015, SKA-TEL-SKO-0000008, SKA Phase 1 System Level 1 Requirements (Rev 6). www.skatelescope.org.
[5] Deboer, D. R., Parsons, et al., "Hydrogen epoch of reionization array (HERA)", Publications of the Astronomical Society of the Pacific, 129(974). doi:10.1088/1538-3873/129/974/045001.
[6] Ewall-Wice et al. "The Hydrogen Epoch of Reionizaiton array dish II. Characterization of spectral structure with electromagnetic simulations and its science implications", Astrophysical Journal, 831(2). doi:10.3847/0004-637X/831/2/196.
[7] N. Fagnoni, E. de Lera Acedo, "The Hydrogen Epoch of Reionization Array (HERA). Improvement of the antenna response with a matching network and scientific impacts", In Proceedings of the 2016 18th International Conference on Electromagnetics in Advanced Applications, ICEAA 2016 (pp. 629-632). doi:10.1109/ICEAA.2016.7731474.
[8] T. D. Carozzi, G. Woan, "A fundamental figure of merit for radio polarimeters," arXiv:0908.2330, 2009.
[9] Saul Kohn et al., "Polarized Foreground Power Spectra from the HERA-19 Commissioning Array,", arXiv:1802.04151v1.
[10] E. de Lera Acedo, N. Drought, B. Wakley, A. Faulkner, "Evolution of SKALA (SKALA-2), the log-periodic array antenna for the SKA-low instrument", in Proc. IEEE International Conference on Electromagnetics in Advance Applications, Turin, September 2015.
[11] De Lera Acedo, E., Trott, C. M., Wayth, R. B., Fagnoni, N., Bernardi, G., Wakley, B., . . . bij de Vaate, J. G. "Spectral performance of SKA log-periodic antennas I: Mitigating spectral artefacts in SKA1-LOW 21 cm cosmology experiments", Monthly Notices of the Royal Astronomical Society, 2017, [1] 469(3), 2662-2671. doi:10.1093/mnras/stx904.

ORCID : https://orcid.org/0000-0001-8530-6989